\documentclass[10pt]{iopart}
\usepackage{iopams}
\usepackage{graphicx}
\usepackage{url,graphics,epsfig}
\usepackage{amssymb}
\usepackage[breaklinks=true,colorlinks=true,linkcolor=blue,urlcolor=blue,citecolor=blue]{hyperref}
\usepackage{cite}

\begin{document}
\jl{2}
%
%
%
\def\etal{{\it et al~}}
\def\newblock{\hskip .11em plus .33em minus .07em}
%
%
%
%
%
\setlength{\arraycolsep}{2.5pt}             

\title[Valence shell photoionization of  W$^{4+}$  ions] {Photoionization of tungsten ions: experiment  and theory for W$^{4+}$}

\author{A M\"{u}ller$^1\footnote[1]{Corresponding author, E-mail: Alfred.Mueller@iamp.physik.uni-giessen.de}$,
		S Schippers$^{1,2}$, J Hellhund$^{1,4}$, A L D Kilcoyne$^3$,\\ R A Phaneuf$^4$
	           and B M McLaughlin$^{5,6}\footnote[2]{Corresponding author, E-mail: bmclaughlin899@btinternet.com}$}

\address{$^1$Institut f\"{u}r Atom- ~und Molek\"{u}lphysik,
                         Justus-Liebig-Universit\"{a}t Gie{\ss}en, 35392 Giessen, Germany}

\address{$^2$I. Physikalisches Institut,
                           Justus-Liebig-Universit\"{a}t Gie{\ss}en, 35392 Giessen, Germany}

\address{$^3$Advanced Light Source, Lawrence Berkeley National Laboratory,
                          Berkeley, California 94720, USA }

\address{$^4$Department of Physics, University of Nevada,
                          Reno, NV 89557, USA}

\address{$^5$Centre for Theoretical Atomic, Molecular and Optical Physics (CTAMOP),
                          School of Mathematics and Physics,
                          Queen's University Belfast, Belfast BT7 1NN, UK}

\address{$^6$Institute for Theoretical Atomic and Molecular Physics,
                          Harvard Smithsonian Center for Astrophysics, MS-14,
                          Cambridge, MA 02138, USA}
%
%
%

\begin{abstract}
Experimental and theoretical results are reported  for single-photon single ionization of  the tungsten ion W$^{4+}$.
Absolute cross sections have been measured  employing the photon-ion merged-beams setup at the Advanced Light Source in Berkeley. Detailed photon-energy scans were performed  at 200~meV bandwidth
in the 40 -- 105~eV range. Theoretical results have been  obtained from a Dirac-Coulomb R-matrix approach employing basis sets of 730 levels for the photoionization of W$^{4+}$. Calculations were carried out for the $4f^{14}5s^2 5p^6 5d^2 \; {^3}{\rm F}_{J}$, $J$=2, ground level and the associated
fine-structure levels with $J$=3  and 4 for the W$^{4+}$ ions.  In addition, cross sections  have been calculated for the metastable levels
$4f^{14}5s^2 5p^6 5d^2 \; {^3}{\rm P}_{0,1,2},{^1}{\rm D}_{2},{^1}{\rm G}_{4},{^1}{\rm S}_{0}$.
Very satisfying agreement of theory and experiment is found for the photoionization cross section of W$^{4+}$ which is remarkable
given the complexity of the electronic structure of tungsten ions in low charge states.
\end{abstract}


\vspace{0.25cm}
\begin{flushleft}
Short title: Valence shell photoionization of  W$^{4+}$ ions\\
\vspace{0.25cm}
\submitto{\jpb: \today, Draft }
\end{flushleft}
%
\maketitle
%
%
%
\section{Introduction}
Tungsten, the element with atomic number $Z=74$, has moved into the focus of controlled nuclear fusion research because of its unique physical and chemical properties which make it the most suitable material for the wall regions of highest particle and heat load in a fusion reactor vessel~\cite{Neu2013}. The downside of tungsten as a high-$Z$ impurity is its extremely high potential for radiative plasma cooling.  Minuscule concentrations of tungsten ions in a fusion plasma prevent ignition. Maximum tolerable relative fractions of tungsten in the plasma are of the order of 2$\times$10$^{-5}$~\cite{Neu2003,Puetterich2010a}.
By plasma-wall interactions tungsten atoms and ions are inevitably released from the surfaces of the vacuum vessel and enter the plasma. Therefore, the understanding of the role of tungsten atoms and ions in a plasma is an essential task of fusion research and development. For the modeling of tungsten plasma impurities and their characteristic line emissions, detailed knowledge about collisional and spectroscopic properties is required. In order to meet some of the most important requirements a dedicated experimental project was initiated several years ago with the goal to provide cross section data and spectroscopic information on tungsten ions exposed to collisions with electrons and photons~\cite{Mueller2015b}. The main topics of this project are electron-ion recombination, photoionization, and electron-impact ionization of tungsten ions. Results on the recombination of W$^{18+}$, W$^{19+}$, and W$^{20+}$ have been published~\cite{Schippers2011b,Spruck2014,Badnell2016,Krantz2014,Krantz2017}. Cross sections and rate coefficients for electron-impact ionization of W$^{17+}$ and W$^{19+}$ have also been made available in the literature~\cite{Rausch2011a,Borovik2016}. The present work adds experimental and theoretical cross sections for single-photon single ionization of W$^{4+}$  to a series of photoionization studies on tungsten ions in low charge states~\cite{Mueller2011a,Mueller2012,Mueller2014c,Mueller2015h,McLaughlin2016a}.

Measurements of collisional properties of tungsten ions have been mainly restricted to the work within the tungsten project mentioned above. Details of the experimental and partly also the theoretical work have been discussed in a recent review~\cite{Mueller2015b} and references to a rapidly growing number of theoretical investigations on tungsten atoms in a wide range of charge states have also been provided there. Previous results on photoabsorption by neutral tungsten atoms have served as benchmarks for ab initio photoionization cross section calculations for W atoms~\cite{Ballance2015a} using the Dirac-Coulomb R-matrix approximation~\cite{Burke2011} which is implemented in the DARC suite of codes~\cite{darc}. DARC uses the complete Dirac Hamiltonian in the inner-region R-matrix calculations. The Coulomb scattering approximation assuming Coulomb wave functions and Coulomb-scattering boundary conditions is used in the outer-region.

Photoemission by tungsten ions is the basis for plasma diagnostics employing high-resolution spectroscopy. Experimental and theoretical studies of this important topic are numerous. An overview of related issues and references to theoretical and experimental work has been provided by Beiersdorfer \etal~\cite{Beiersdorfer2015}.

As a rule, particle densities in magnetically confined fusion plasmas are too low to support significant absorption of photons emitted by the plasma constituents. Therefore, the fusion-motivated relevance of photoabsorption by tungsten atoms and ions is limited. Photonic interactions are only expected to play a role in the plasma edge region where dust grains may shield radiation emitted from the plasma core~\cite{Brown2014}.
Photoionization of tungsten atoms and ions is of plasma-related interest nevertheless because
it can provide details about spectroscopic properties of tungsten which are needed for plasma diagnostics  and, as time-reversed photorecombination, photoionization may help to
better understand one of the most important atomic collision processes in a fusion plasma, namely, electron-ion recombination.

Beside its importance in the field of controlled nuclear fusion, tungsten is of considerable interest as a prototypical high-$Z$ element that provides ample opportunities for studying many-electron effects on the structure and collisional properties of atoms and ions. The present Dirac-Coulomb R-matrix approximation is one of the most advanced theoretical tools to obtain information about electron-ion and photon-ion interactions in general and is suitable, in particular, for the treatment of atoms and ions characterized by a complex electronic structure. Studies on photoionization of tungsten atoms and ions with their complex electronic structure featuring open $d$ and $f$ shells  and comparison of experimental and theoretical results can provide benchmarks and guidance for future theoretical work on electron-ion interaction processes. The direct and resonant photoionization  processes  occurring for the Yb-like W$^{4+}(4f^{14} 5s^2 5p^6 5d^2)$ ions in the present energy range comprise removal or excitation of either a $4f$, $5s$, $5p$ or $5d$ electron. For the theoretical description of W$^{4+}$ ions undergoing the photoionization process, suitable target wave functions have  to be constructed that allow for promotions
of electrons from the $4f, 5s, 5p,$ and $5d$ subshells to all contributing excited states.
This is extremely challenging for low-charged tungsten ions. For the tungsten ions in higher charge states, such as  W$^{4+}$,
it is still challenging but expected to be a slightly less daunting task, due primarily to the increased effect of the Coulomb charge of the target and the slight reduction in the R-matrix box size.

In this paper we report on experimental and theoretical cross sections for single photoionization of W$^{4+}$ ions.  This work
provides the next step in our present series of investigations on single-photon single ionization of tungsten atoms/ions in low charge states.
The layout of this paper is  as follows. Section 2 details the experimental procedure.
Section 3 presents a brief outline of the theoretical work. Section 4 presents a discussion of the
results obtained from both the experimental and theoretical methods.
Finally in section 5 conclusions are drawn from the present investigation.
%
%
%
%
%

\section{Experiment}\label{sec:exp}

The experiments on photoionization of W$^{4+}$ ions made use of the
Ion-Photon Beam (IPB) endstation of beamline 10.0.1.2 at the
Advanced Light Source (ALS) in Berkeley, California, USA. For the measurement of absolute cross sections the merged-beams technique~\cite{Phaneuf1999} was employed. Recently, experimental methods for studying photoabsorption by ions and typical results of such experiments have been reviewed by M\"{u}ller \etal~\cite{Mueller2015b} and Schippers \etal~\cite{Schippers2016}. The general layout
of the  IPB setup and the associated experimental procedures have
been described by Covington \etal~\cite{Covington2002a}. Since the IPB endstation was first implemented nearly two decades ago significant technological improvements have been made. The most recent account of measurements at the IPB has been published by Macaluso \etal~\cite{Macaluso2016}. A detailed description of the methodology used for the photoionization of tungsten ions has been provided in our publication on the results for W$^+$ ions~\cite{Mueller2015h}.

Here, only an  overview of the experiment is given. Aspects specific to the present measurements are discussed in more detail.
For the preparation of beams of tungsten ions W(CO)$_6$ vapour was leaked via a needle valve into the plasma chamber of a 10~GHz electron-cyclotron-resonance (ECR) ion source~\cite{Trassl1997a}. A steady discharge was maintained by adding Ar or Xe as a support gas.  A mixture of ions produced from the support gas, the W(CO)$_6$ vapour and other materials present in the plasma chamber was accelerated by a voltage of typically $U_{\rm acc}$ = 6~kV and an ion beam was formed by a suitable set of electrostatic focusing elements. By a subsequent 60$^o$ dipole magnet the desired beam component was selected and directed towards an electrostatic spherical 90$^o$ deflector (the merger) which deflected the ion beam onto the photon beam axis. A mass-over-charge spectrum of ions produced in this manner is provided in the overview~\cite{Mueller2015b} of  the tungsten project initiated at Giessen University. In addition to many exotic fragment species produced from W(CO)$_6$ molecules atomic tungsten ions, W$^{q+}$, in a wide range of charge states up to $q=19$ were also obtained. Ion currents of collimated beams of isotope-resolved $^{186}$W$^{4+}$ ions employed in the present experiments were as large as 30~nA.

Behind the merger the selected W$^{4+}$ ion beam passed the  interaction region which was essentially an electrically isolated drift tube of about 29~cm length. For the measurement of absolute cross sections the interaction region was set to a potential of up to $U_{\rm D}$ = 1~kV in order to tag product ions (electrical charge $5 e$) from within the interaction region by their final energy $4 e U_{\rm acc} + e U_{\rm D}$. A 45$^o$ dipole magnet, the demerger, separated the W$^{4+}$ parent ion beam from the W$^{5+}$ products which were deflected out of plane by a spherical 90$^o$ deflector and directed towards a single-particle detector with almost 100\% efficiency~\cite{Fricke1980a,Rinn1982}. The energy difference $e U_{\rm D}$ between product ions from outside and from inside the interaction region was sufficient for complete separation of the two components by the demerger magnet. The primary ion beam current was collected by a large Faraday cup inside the demerger magnet. Separation of photoionized ions from background  was accomplished by mechanically chopping the photon beam and by phase-sensitive recording of detector pulses.

Significant contaminations of the $^{186}$W$^{4+}$ beams with molecular ions  featuring the identical mass-over-charge ratio (46.5) can safely be ruled out considering their composition of C and O atoms and the effect that they would have on the observed distribution of tungsten isotopes in the mass spectrum which followed the natural abundances. Moreover the setting of the second magnet to the W$^{5+}$ product channel precludes the observation of single ionization of any primary ion other than W$^{4+}$.

For absolute cross section measurements the beam overlap factor was determined by scanning x- and y-profiles of the ion beam and the counter-propagating photon beam  at three positions along the z-axis in the middle and at the front and rear ends of the interaction region. By energy-tagging the product ions the length of the interaction region was defined as the length of the isolated drift tube. The error budget of the absolute cross sections obtained by this procedure has been discussed previously~\cite{Mueller2014b} and a total systematic uncertainty of 19\% was estimated. This uncertainty does not include problems with the purity of the two merging beams. In a thorough investigation described by M\"{u}ller \etal in the context of photoionization of W$^+$ ions~\cite{Mueller2015h} energy-dependent fractions of higher-order radiation in the photon beam up to the sixth order could be detected and quantified. A procedure for correcting measured apparent cross sections was developed both with respect to correct normalization of photoionization signal to the photon flux and removal of surplus signal arising from photoionization at the higher energies $nE_\gamma$ of the $n$th order radiation fractions. We note that even-order contributions could be only partly suppressed by tightly closing the baffles behind the monochromator and by thus losing most of the flux of the photon beam. Corrections were made for the second- and third-order contaminations of the photon beam neglecting the smaller effects of radiation orders $n \geq 4$. The uncertainties of this procedure were added to the total possible error of the measured absolute cross sections as described previously~\cite{Mueller2015h}. The uncertainty of the energy axis in the present experiments is estimated to be $\pm 200$~meV.

A further problem in experiments employing beams of ions with a complex electronic structure is the possible presence of ions in long-lived excited states. This problem has been discussed previously~\cite{Mueller2015b,Mueller2015h} and is particularly relevant for tungsten ion beams. It is illustrated in the context of photoionization of W$^{4+}$ ions by the following considerations.
The minimum electron energy required for efficient production of W$^{q+}$ ions in a continuously operating ion source is determined by the ionization threshold of the associated W$^{(q-1)+}$ ions. The minimum energy required for producing W$^{4+}$ from W$^{3+}$ is $38.2 \pm 0.4$~eV~\cite{NIST2015}. There must be a sufficient density of electrons in the discharge carrying about three times that energy~\cite{Mueller2008a} to facilitate efficient ionization of W$^{3+}$. The charge-state spectrum of tungsten ions measured with the ion-source settings for the optimum production of W$^{4+}$ ions even showed production of charge states up to W$^{19+}$ requiring electron energies well above the ionization potential of about 460~eV of W$^{18+}$. Thus, the hot component of the electron energy distribution~\cite{Gumberidze2010}  has to have a minimum temperature corresponding to $kT \gtrapprox 500$~eV, because otherwise, the high charge states would not have been produced. The weight of level $\langle i \rangle$ with excitation energy $E_i$ in a thermodynamical equilibrium is  proportional to $g_i exp(-E_i/(kT))$, where $g_i$ is the statistical weight of level $\langle i \rangle$. Here the excitation energies are below the ionization energy of 38.2~eV while $kT$ is greater than 500~eV. Hence, the exponent is greater than $exp(-38.2/500)=0.93$, close to 1. Although there is not really a thermodynamical equilibrium in the source, this argument should be valid for the hot electron component in the source and for all excited states of W$^{4+}$.  Under such conditions, every excited state of  W$^{4+}$ ions can be populated in the ion source with its statistical weight.

The ions, once produced, are confined in the source plasma for millisecond times~\cite{Hitz2000} before they drift out and are extracted and accelerated. Flight times of the accelerated tungsten ions of interest were of the order of 20~${\mu}$s between the ion source and the photon-ion interaction region. Most of the excited levels have decayed after such long times. However, metastable ions can survive and contribute to the photoionization signal. Obvious candidates for metastable excited states in W$^{4+}$ are associated with the $5p^65d^2$ ground-state configuration. All the
levels within one given configuration are long lived since electric dipole transitions between any two states  within the configuration are strictly forbidden. Therefore, the parent ion beam must be expected to contain ions in all levels associated with the $5p^65d^2$ configuration. The ground level of the Yb-like atomic tungsten ion, W$^{4+}$,  is $5p^65d^2~^3{\rm F}_{2}$. In addition, the levels $5p^65d^2~^3{\rm F_{3,4}}$, $5p^65d^2~^3{\rm P_{0,1,2}}$, $5p^65d^2~^1{\rm G_4}$, $5p^65d^2~^1{\rm D_2}$ and $5p^65d^2~^1{\rm S_0}$ have to be considered when comparing experimental results with theory.

%
%
%
%
%
\section{Theory}\label{sec:Theory}

State-of-the art theoretical methods were employed to calculate photoionization cross sections for comparison with experiment. The calculations utilized a parallel version~\cite{Ballance2006,Fivet2012,McLaughlin2012a,McLaughlin2012b} of the DARC package of computer codes~\cite{Norrington1987,Wijesundera1991,darc} designed to investigate interactions of electrons and photons with atoms and their ions. Highly correlated wavefunctions that incorporate relativistic effects were constructed for the hundreds of levels that were considered and scattering calculations were performed for thousands of channels. Such extensive calculations require access to parallel high-performance computer architectures and are currently being performed on a number of platforms worldwide~\cite{McLaughlin2016a,McLaughlin2015a,McLaughlin2015b,McLaughlin2016c,McLaughlin2017b}.
Recently, DARC calculations on photoionization cross sections
were carried out for Se$^{+}$\cite{McLaughlin2012b}, Se$^{2+}$ \cite{Macaluso2015},
Kr$^{+}$ \cite{Hinojosa2012,McLaughlin2012a}, Xe$^{+}$\cite{McLaughlin2012a},
Xe$^{7+}$ \cite{Mueller2014b},
2p$^{-1}$ inner-shell studies on Si$^{+}$ \cite{Kennedy2014}, Ar$^{+}$ \cite{Tyndall2016a}, and Co$^+$ \cite{Tyndall2016b},
valence-shell studies on neutral sulphur \cite{Barthel2015}, sulphur-like chlorine, Cl$^+$  \cite{McLaughlin2017c},
tungsten and its ions; W \cite{Ballance2015a}, W$^{+}$ \cite{Mueller2015h},
W$^{2+}$ and W$^{3+}$ \cite{McLaughlin2016a}.
All of these cross section calculations using the DARC codes
showed suitable agreement with the measurements made at the Advanced Light Source, BESSY II and SOLEIL radiation facilities.
In the present work we concentrate our efforts on
photoionization cross section calculations for the
W$^{4+}$ ion to accomplish the next step in our investigations on
low charged states of atomic tungsten ions.

%
%
\begin{table}
\footnotesize
\caption{Comparison of the NIST~\cite{NIST2015} tabulated data with the present theoretical energies
	    obtained by using the GRASP code for Tm-like W (W$^{5+}$) ions. Relative energies with respect to the ground state
               are given  in Rydbergs (Ry).  A sample of the nine  lowest NIST levels of the residual W$^{5+}$ ion are
               compared with progressively larger GRASP calculations,  (52-level, 385-level,  669-level and 730-level approximations)
               and previous relativistic-configuration-interaction (RCI) and multiconfiguration-Hartree-Fock (MCDHF) calculations by Enzonga Yoca \etal \cite{EnzongaYoca2012a} using the extended optimal level (EOL) option.}
\label{tab1}.
\begin{tabular}{lccccccccc}
\br
Level         	&  State	     	&  Term		& NIST  		 &GRASP		&GRASP 	    	 &GRASP      &GRASP       	&RCI		&MCDHF\\
		&			&			& Energy$^{a}$  &Energy$^{b}$  &Energy$^{c}$ &Energy$^{d}$ &Energy$^{e}$ &Energy$^{f}$&Energy$^{g}$ \\	
		&			&			& (Ry)		 	&(Ry)			&(Ry) 			& (Ry) 	&(Ry)  	& (Ry)		& (Ry)\\	
\mr
 1  		& $5d$ 	&  $\rm^2D_{3/2}$ 	& 0.00000		&0.00000		&0.00000		&0.00000 	&0.00000	&0.00000	&0.00000\\
 2  		& 		&  $\rm^2D_{5/2}$ 	& 0.07937		&0.07351		&0.07492		&0.07170	&0.07170	&0.07572	&0.07297\\
 \\
 3  		& $6s$   & $\rm^2S_{1/2}$   		& 0.72383		&0.72210		&0.68176		&0.65632	&0.65847	&0.70359	&0.73271\\	
\\
 4  		& $6p$ &  $\rm^2P^o_{1/2}$		 &1.34460		&1.32412		&1.29278		&1.25928	&1.25810	&1.31193	&1.34326\\
 5  		&          &  $\rm^2P^o_{3/2}$ 		&1.50393		&1.47783		&1.44784		&1.40929	&1.41059	&1.46578	&1.49284\\
\\
6		& $5f$  &  $\rm^2F^o_{5/2}$ 		&2.38461		&2.76920		&2.77331	  	&2.30579	&2.30385	&2.35477	&2.39873\\
7 		&         &  $\rm^2F^o_{7/2}$ 		&2.39144		&2.82627		&2.83061		&2.30755	&2.30674	&2.36070	&2.40059\\
\\
8  		& $6d$ & $\rm^2D_{3/2}$     		&2.38474		&2.33822		&2.32703		&2.28443	&2.28678	&2.34464	&2.35272\\
9 		&          &  $\rm^2D_{5/2}$    		&2.40950		&2.36227		&2.35156	   	&2.30844	&2.31074	&2.36891	&2.37653\\
\mr
\end{tabular}
\\
\begin{flushleft}
$^{a}$NIST Atomic Spectra Database, tabulations \cite{NIST2015}.\\
$^{b}$GRASP, present 52-level approximation.\\
$^{c}$GRASP, present 385-level approximation.\\
$^{d}$GRASP, present 669-level approximation.\\
$^{e}$GRASP, present 730-level approximation.\\
$^{f}$GRASP2k, RCI approximation~\cite{EnzongaYoca2012a}.\\
$^{g}$GRASP2k, MCDHF-EOL approximation~\cite{EnzongaYoca2012a}. \ \\
\end{flushleft}
\end{table}
To investigate single photoionization of the  W$^{4+}$ ion, we began with
a very simple 52-level  approximation arising from the 5 configurations
$4f^{14}5s^25p^65d$, 	$4f^{14}5s^25p^55d^2$,
 $4f^{14}5s^25p^66s$,  $4f^{14}5s^25p^66p$ and $4f^{14}5s^25p^66d$
for the wavefunctions of the W$^{5+}$ product ion, with only the $5p$ shell being opened,
allowing $5d  \rightarrow 6\ell$ and $5p \rightarrow 5d$ electron excitations.
This model was extended with the addition of 10 further configurations
by opening the $5s$ shell allowing $5s \rightarrow n\ell$ and
$5d \rightarrow n\ell$ excitations together with
$5p \rightarrow n\ell$ promotions, $n$=5 and 6,
resulting in the selected electron configurations
$4f^{14}5s5p^65d^2$,  $4f^{14}5s5p^65d6s$,  $4f^{14}5s5p^65d6p$, $4f^{14}5s5p^65d6d$,
$4f^{14}5s^25p^5 5d 6s$,  $4f^{14}5s^25p^5 5d 6p$, $4f^{14}5s^25p^5 5d 6d$,
$4f^{14}5s^25p^56s^2$, $4f^{14}5s^25p^56p^2$ and $4f^{14}5s^25p^56d^2$. This 15-configuration model
gave 385 levels for the  W$^{5+}$ product-ion wavefunctions.

Next we  investigated  a model where we opened the $5p$ and $4f$ shells
and included $4f \rightarrow 5d$, $5p \rightarrow  6\ell$, $5d \rightarrow 6\ell$ promotions,  and  selected
$5p^2 \rightarrow 6 \ell^2$ double electron promotions.
The configurations included were
$4f^{14}5s^25p^65d$,
$4f^{14}5s^25p^66s$,  $4f^{14}5s^25p^66p$,  $4f^{14}5s^25p^66d$,
$4f^{13}5s^25p^65d^2$, $4f^{14}5s^25p^55d6s$,  $4f^{14}5s^25p^55d6p$, $4f^{14}5s^25p^55d6d$,
$4f^{14}5s^25p^45d6s^2$, and $4f^{14}5s^25p^45d6p^2$.
This 10-configuration model yielded
a total of 669 levels for the W$^{5+}$ product ion and improved the residual ion energies
with the opening of the $4f$ shell.  The  $4f^{14}5s^25p^45d6d^2$  configuration
was neglected as it would have added an additional 921 levels to the close-coupling
collision model and therefore made calculations untractable.

A final model was investigated where we opened the $5s$, $5p$ and $4f$ shells and included $5s \rightarrow 5d$, $4f  \rightarrow  5d$,
$5p  \rightarrow 5d, 6\ell$ and $5d \rightarrow 6\ell$ single as well as
$5p^2 \rightarrow 6 \ell^2$ double electron promotions.
The configurations included were:
$4f^{14}5s^25p^65d$,
$4f^{14}5s^25p^66s$,  $4f^{14}5s^25p^66p$,  $4f^{14}5s^25p^66d$,
$4f^{14}5s5p^65d^2$, $4f^{13}5s^25p^65d^2$,
$4f^{14}5s^25p^55d^2$, $4f^{14}5s^25p^55d6s$, $4f^{14}5s^25p^55d6p$, $4f^{14}5s^25p^55d6d$,
$4f^{14}5s^25p^45d6s^2$, and $4f^{14}5s^25p^45d6p^2$.
The  $4f^{14}5s^25p^45d6d^2$  configuration
was once again neglected for the reason  stated above.
This twelve-configuration model yielded a total of 730 levels for the W$^{5+}$ product ion.

Table 1 gives a comparison of the W$^{5+}$ level-energy approximations obtained from the GRASP code
with the tabulated values from the  NIST~\cite{NIST2015} database
and with previous large-scale calculations~\cite{EnzongaYoca2012a}.
Note the difficulty of accurately describing the energy levels in the various approximations.
Naturally, much larger basis-state expansions than those considered here would probably bring the slowly converging energies  into better agreement with experiment.
Such larger-scale expansions would be prohibitive to performing photo-ionization cross section calculations. The present approximation has to be viewed as a compromise between a suitable representation of this tungsten ion structure and the feasibility of performing the photoionization computations at the
present technical limit of {\em ab initio} close-coupling treatment.

Photoionization cross section calculations for the 730-level approximation
 were then carried out in  the Dirac-Coulomb approximation using the
DARC codes \cite{McLaughlin2012a,McLaughlin2012b} for photon
energies from the ionization thresholds up to 150~eV.
The R-matrix boundary radius of 10.88 Bohr radii  was sufficient to
envelop the radial extent of all the n=6 atomic orbitals
of the  W$^{5+}$ product ion. A basis of 12 continuum orbitals was
sufficient to span the  photon energy range chosen
for the calculations.   Photoionization cross-section calculations were
performed for all the terms of the $4f^{14}5s^2 5p^65d^2$ configuration,
namely; $4f^{14}5s^2 5p^6 5d^2 \; {^3}{\rm F}_{2,3,4}$,  $4f^{14}5s^2 5p^6 5d^2 \; {^3}{\rm P^{o}}_{0,1,2}$,
 $4f^{14}5s^2 5p^6 5d^2 \; {^1}{\rm D}_{2}$,  $4f^{14}5s^2 5p^6 5d^2 \; {^1}{\rm G}_{4}$, and
 $4f^{14}5s^2 5p^6 5d^2 \; {^1}{\rm S}_{0}$.
For the $4f^{14}5s^2 5p^6 5d^2 \; {^3}{\rm F}_{J}$, $J$=2, ground state,
since dipole selection rules apply,  total ground-state
photoionization cross sections require only
 the  bound-free dipole matrices,
$J^{\pi}=2^{e} \rightarrow J^{\pi}=1^{\circ},2^{\circ},3^{\circ}$.
Whereas for the excited metastable states associated with $5d^2$ configuration then,
$J^{\pi}=3^{e} \rightarrow J^{\pi}=2^{\circ},3^{\circ},5^{\circ}$,
 $J^{\pi}=4^{e} \rightarrow J^{\pi}=3^{\circ},4^{\circ},5^{\circ}$,
 $J^{\pi}=1^{e} \rightarrow J^{\pi}=0^{\circ},1^{\circ},2^{\circ}$  and
 $J^{\pi}=0^{e} \rightarrow J^{\pi}=1^{\circ}$ are all necessary.

For the ground and metastable initial states of the W$^{4+}$  ion
studied here, the outer region electron-ion collision problem was
solved (in the resonance region below and  between all thresholds)
using a  fine energy mesh of $\approx$ 0.170 meV for all levels investigated.
The present close-coupling calculations for W$^{4+}$ ions resulted in
approximately 4,500 channels with Hamiltonians
and dipole matrices in excess of 60,000 in size which are solved in the Dirac-Coulomb approximation~\cite{Burke2011} to obtain
the photoionization cross-sections..
The $jj$-coupled Hamiltonian diagonal matrices were adjusted
so that the theoretical term energies matched the recommended
NIST values~\cite{NIST2015}. The energy shifts between the 730-level approximation and the NIST values were between -0.10 and -1.59~eV with an average of -1.0785~eV. The average shift was used in cases where no NIST energies were available. This energy adjustment ensures better positioning of resonances relative to all thresholds included in the calculation \cite{McLaughlin2012a,McLaughlin2012b}. We note that the length and velocity forms always agree with one another at the few-percent level or better in such large-scale calculations. In the present case the cross sections were  calculated in the length form.

%
%
%
\begin{figure*}
\centering
\includegraphics[width=13cm]{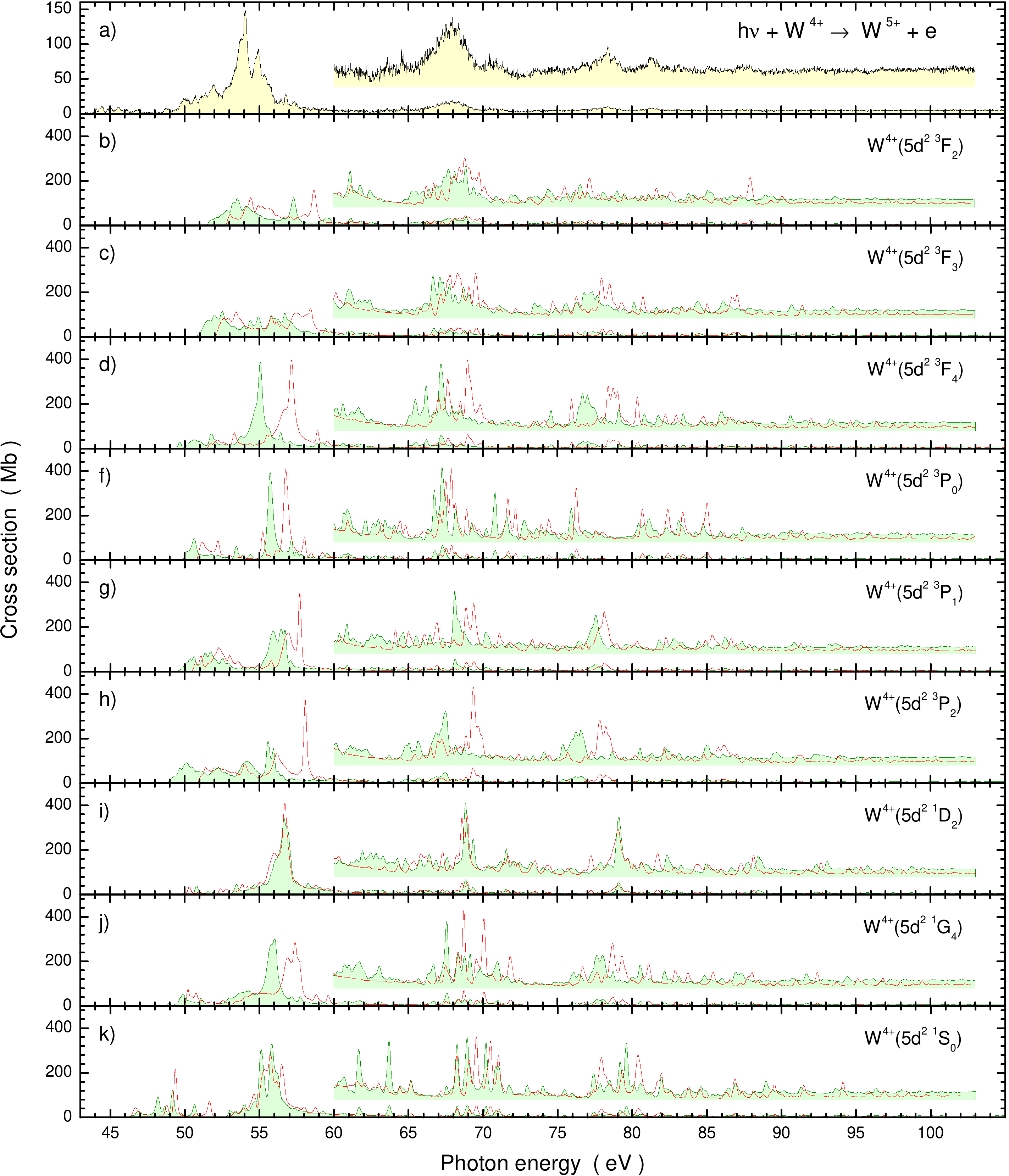}
\caption{\label{Fig:W4over}(Colour online) Overview of the present experimental (panel (a)) and theoretical (panels (b) through (k)) photoionization cross sections for W$^{4+}$ ions as a function of photon energy. The calculated cross sections were convoluted
with 200-meV FWHM Gaussians such as to model the experiment. In each panel, the 60--93-eV cross-section functions are shown multiplied by a factor of 5 and vertically offset by 40~Mb. The theoretical cross sections for all nine levels within the $4f^{14} 5s^2 5p^6 5d^2$ ground-state configuration are presented with each panel showing the spectroscopic notation of the associated level. The (red) solid line without shading is the result of the 52-level model, the (olive) solid line with light (green) shading represents the result of the more sophisticated 730-level approximation.}
\end{figure*}

\section{Results and Discussions}\label{sec:Results}

In figure \ref{Fig:W4over} we present the experimental and theoretical photoionization cross-section results for the W$^{4+}$ ion. The measurements were carried out at a constant energy resolution of 200~meV. The experimental cross section in panel (a) is represented by the measured energy-scan results normalized to a number of absolute data points. The dominant feature in the spectrum is a broad jagged peak structure at energies between about 49 and 60~eV. Smaller but similarly complex peak structures follow at energies around 68, 78, and 81~eV. The origin of the different peaks and their assignment to certain excitation processes is not immediately obvious. A problem is in the very large number of possible transitions in the investigated energy range. Excitation $5p \to 6s$  from the ground configuration $4f^{14} 5s^2 5p^6 5d^2$ to the $4f^{14} 5s^2 5p^5 5d^2 6s$ excited configuration involves 400 dipole transitions spread out over an energy range from about 39 to 65 eV. If the final subshell is $6p$ instead of $6s$ the number of possible dipole transitions goes up to 1546 and the energy range is from about 59 to 85 eV. The situation becomes even more unclear  when the $4f$ subshell is opened by a one-electron excitation with thousands of possible transitions.

The energy ranges provided above are from fine-structure--resolved calculations with the Cowan code~\cite{Cowan1981}. The calculation of configuration-averaged energies and associated oscillator strengths gives a clearer picture. Excitations $5p \to ns$ and $5p \to nd$ as well as $4f \to nd$ and $4f \to ng$ with $n=6,7,8,9$ are within the range of the peak features, while  $5s$ excitations occur at higher energies. The largest oscillator strengths are associated with $5p \to ns$ transitions, followed by $5p \to nd$ transitions. Thus, the calculations suggest that the main resonance group at 54~eV is mainly composed of $5p \to 6s$ excitations with contributions of $5p \to 5d$. The peak at 68 eV might be associated with $5p \to 7s$ and $5p \to 6d$ transitions and the peak features at higher energies may originate from $5p \to nd/(n+1)s$ excitations with $n=7,8,...$.

Theoretical cross sections obtained from the DARC calculations were convoluted with 200-meV FWHM Gaussians to simulate the experimental photon-energy resolution. Results from the 52-level approximation are shown by the (red) solid lines without shading while the more sophisticated large-scale 730-level calculations are represented by the (olive) solid lines with light (green) shading. Panel (b) provides the theoretical cross sections for the W$^{4+}$ ion in the $5d^2~^3{\rm F}_2$ ground level, panels (c) through (k) show the calculated cross sections for the initial metastable levels $5d^2~^3{\rm F}_3$, $5d^2~^3{\rm F}_4$, $5d^2~^3{\rm P}_0$, $5d^2~^3{\rm P}_1$, $5d^2~^3{\rm P}_2$, $5d^2~^1{\rm D}_2$, $5d^2~^1{\rm G}_4$, and $5d^2~^1{\rm S}_0$, respectively.

The resonance strengths contained in the cross section from the 52-level approximation are not very much different from those resulting from the 730-level model. The main effect of the larger basis set with the greatly enhanced large-scale configuration-interaction calculation is a redistribution of resonance strengths within the groups of resonances and in most cases a shift of the groups towards lower energies. This shift results in better agreement of the theoretical with the experimental peak energies.

\begin{figure}
\centering
\includegraphics[width=12cm]{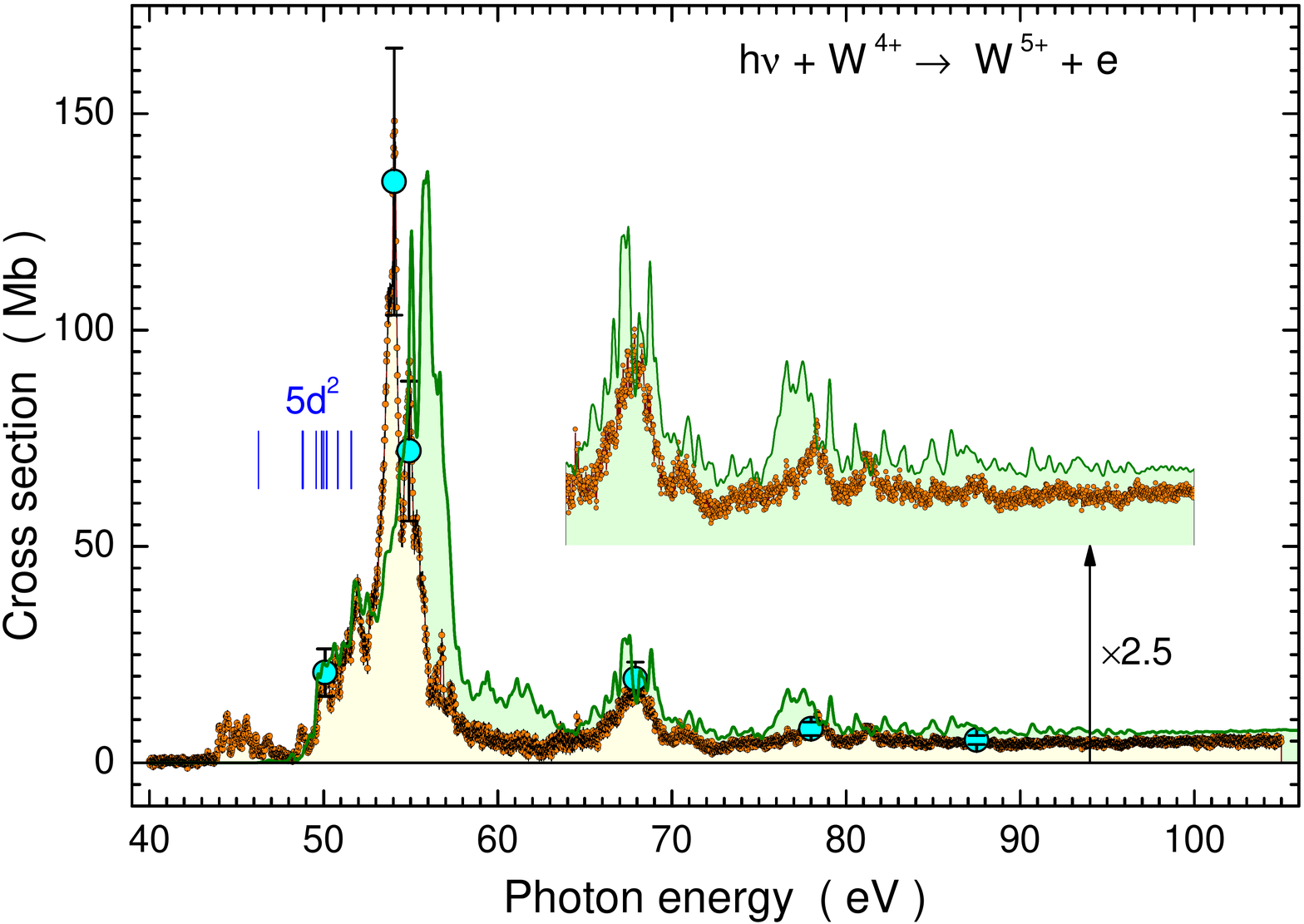}
\caption{\label{Fig:W4comp}(Colour online) Comparison of the present experimental and theoretical photoionization cross sections for W$^{4+}$ ions. The statistically weighted sum of the results for all levels within the $4f^{14} 5s^2 5p^6 5d^2$ configuration, i.e., the configuration-averaged cross section, obtained with the 730-level approximation was convoluted with a 200-meV FWHM Gaussian to simulate the experimental energy resolution. The cross sections at energies beyond 64~eV were multiplied with a factor of 2.5 and displayed separately with a vertical offset of 50~Mb. The experimental energy-scan data are displayed as small circles with (orange) shading. The statistical error bars are provided as vertical black bars which can hardly be seen because they are of the size of the data points. Absolute cross sections are shown as larger circles with (cyan) shading together with their total uncertainties. The theoretical result is represented by the (olive) solid line with light (green) shading. The ionization thresholds of all levels within the $4f^{14} 5s^2 5p^6 5d^2$ ground-state configuration are given by the vertical (blue) lines. They were taken from the NIST compilation~\cite{NIST2015}.
					}
\end{figure}

As discussed in section~\ref{sec:exp} the ground-state configuration $4f^{14} 5s^2 5p^6 5d^2$ of W$^{4+}$ is characterized by two electrons in the $5d$ subshell while all other subshells below the outermost valence shell are completely filled. This configuration supports nine fine-structure levels, all with identical parities. Hence, electric dipole transitions between any of these levels are forbidden and accordingly, all these levels have long lifetimes. Since the maximum excitation energy within the $5d^2$ configuration is only about 5~eV, while the temperature of the hot-electron component in the ion-source plasma is higher by orders of magnitude, it is reasonable to assume that all terms and levels are statistically populated. Statistical population of fine-structure levels within a given term has been verified in previous high-resolution photoionization experiments on berylliumlike ions~\cite{Mueller2010b}. Hence, for a meaningful comparison with the experiment, the theoretical results for the individual fine-structure levels have been added with their statistical weights relative to the total weight (24) of the 5d$^2$ configuration. The resulting configuration-averaged, convoluted theoretical cross section obtained from the 730-level
approximation is compared with the experimental data in figure~\ref{Fig:W4comp}.

%
%
%

%
%
%
\begin{figure}
\centering
\includegraphics[width=12cm]{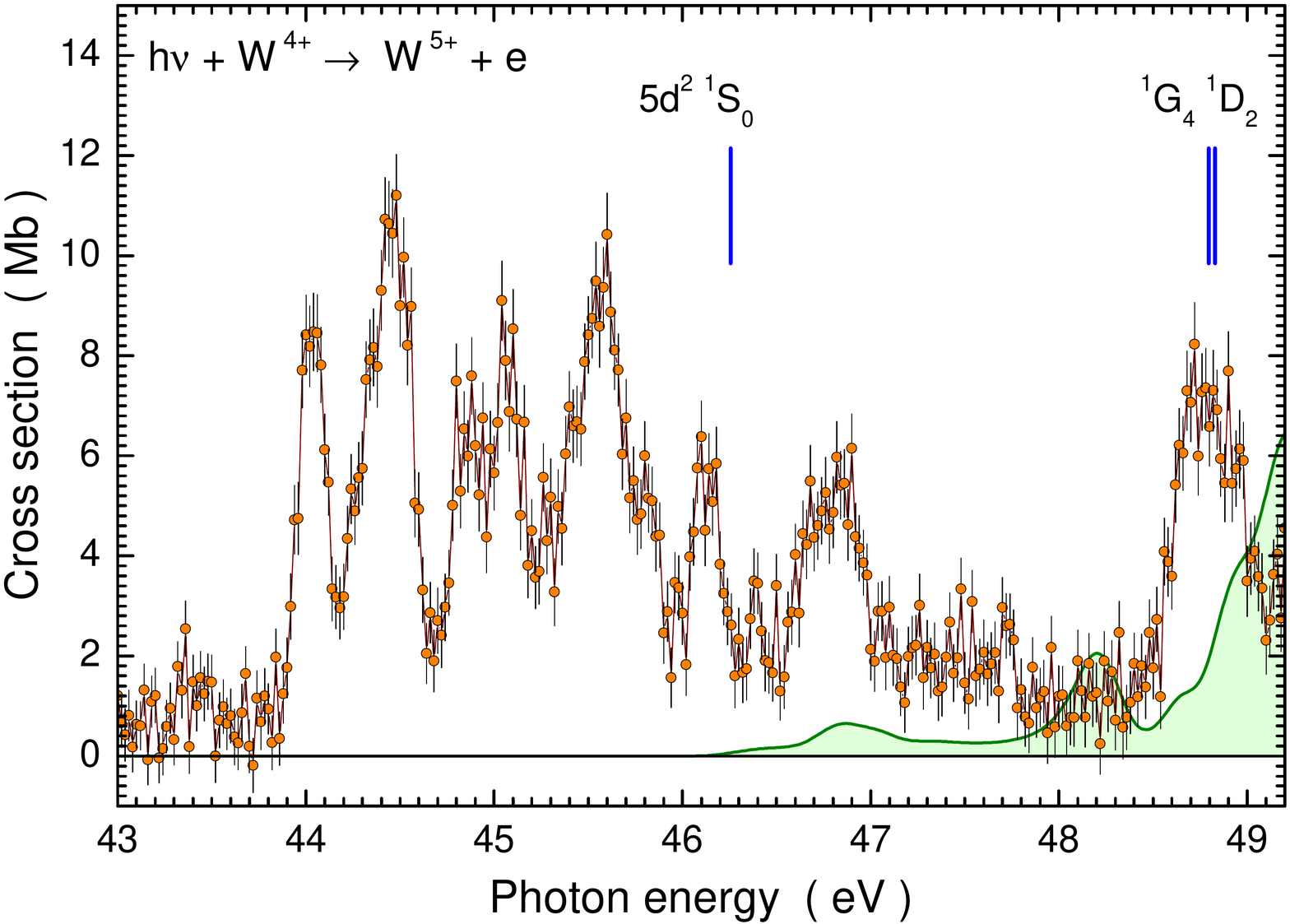}
\caption{\label{Fig:W4lowE}(Colour online) Blow-up of the low-energy cross-section region from the preceding figure. For explanations see the caption of figure~\ref{Fig:W4comp}. The onsets of ionization arising from the three highest levels within the $4f^{14} 5s^2 5p^6 5d^2$ ground-state configuration are identified in the figure. }
\end{figure}

The absolute size of the dominant experimental cross section features is very well reproduced by the large-scale calculation. There are only slight non-systematic shifts of at most $\pm 2$~eV between the theoretical and experimental peak energies. The large basis set chosen for this calculation is obviously sufficient to produce very satisfying results for the photoionization of W$^{4+}$ ions given the complexity of the problem in which several subshells with high angular momenta are involved.

The vertical (blue) bars in figure~\ref{Fig:W4comp} show the ionization thresholds for the $5d^2$ fine-structure levels obtained from the NIST tables~\cite{NIST2015}. A non-vanishing cross section below the ground-state ionization threshold indicates the presence of metastable ions in the parent ion beam. The contributions arising from the long-lived excited states within the $5d^2$ configuration are already included in the theory curve with their statistical weights. Nevertheless, there is still a non-zero cross section at energies below the ionization potential of the most highly excited $5d^2$ fine-structure component, the $5d^2~^1{\rm S}_0$ level.

In order to obtain a clear picture of the low-energy cross-section contributions, a blow-up of the threshold region is shown in figure~\ref{Fig:W4lowE}. At energies below the steep cross-section onset at about 50~eV there is a group of small but distinct resonance features. Some of these features are above the  $5d^2~^1{\rm S}_0$ ionization threshold and therefore might be included in the calculations for the $5d^2$ fine-structure levels. The comparison of the large-scale 730-level calculation with the experimental results shows, however, that the $5d^2~^1{\rm S}_0$ contribution to the cross section is too small to explain the peak features. Clearly, the calculated cross section vanishes at energies below the $5d^2~^1{\rm S}_0$ ionization threshold indicated by the (blue) bar at about 46.3~eV while the experimental cross section continues to stay up until a sharp drop-off is reached at about 43.9~eV. Given the energy shifts of the dominant theoretical from the experimental peak structures at about 55~eV one might suspect that there is also an energy shift of theory in the energy range of figure~\ref{Fig:W4lowE}. The fact that the onset of the theoretical cross section matches the NIST ionization threshold of the $5d^2~^1{\rm S}_0$ level indicates that there is almost no room for such a shift of the theoretical cross section within the energy range of figure~\ref{Fig:W4lowE}. Hence, one has to conclude that additional metastable levels beyond the highest $5d^2$ fine-structure levels must have been present in the parent ion beam of the experiment.

Possible candidates for more highly excited metastable levels can be found within the $5p$- and $4f$-subshell--excited $4f^{14} 5s^2 5p^5 5d^3$ and $4f^{13} 5s^2 5p^6 5d^3$ configurations where quintet states with more than 30~eV excitation energy can be formed. Treating all the 110 + 206 = 316 levels within those two configurations goes beyond presently available computing resources and is also beyond the scope of the present paper. While the peaks at energies below the $5d^2~^1{\rm S}_0$ ionization threshold might be attributed to contributions arising from levels within the inner-subshell excited configurations the size of the apparent cross sections at energies below 46~eV is at most one fifteenth of the cross section maximum at 54~eV. Therefore, the effect on the present W$^{4+}$  photoionization cross section  arising from metastable  levels within the $4f^{13} 5s^2 5p^6 5d^3$ configuration may be expected to be small. Some of the deviations between theory and experiment may have to be attributed, though, to the neglect of more highly excited metastable levels beyond the $5d^2$ ground-state configuration in the parent ion beam.

The DARC calculations for W$^{q+}$ with $q = 0, 1, 2, 3$, and $4$ show increasingly better agreement with experiments along the sequence of increasing charge states. This was expected because the physics of more highly charged ions becomes simpler in that the electron-nucleus interactions become more prominent relative to the electron-electron interactions which are difficult to treat. The comparison of theoretical and experimental cross sections in figure~\ref{Fig:W4comp} with the good agreement observed here provides a very good example supporting the expectation. It would be interesting to see how well other theoretical methods can reproduce the experimental results. A statistical theory based on the concept of quantum many-body chaos has been suggested for the treatment of atomic processes involving interactions of electrons and photons with complex many-electron atoms or ions (\cite{Flambaum2015}. The application of this approach to recombination of tungsten ions with an open $4f$ shell such as  W$^{20+}$~\cite{Berengut2015} provided very good overall agreement with the experiment~\cite{Schippers2011b}. The quantum-chaos theory may turn out to be also suitable for the present problem of photoionization of W$^{q+}$ ions in low charge states $q$.

\section{Summary and Conclusions}\label{sec:Conclusions}
Absolute experimental and theoretical cross sections are presented for single photoionization of W$^{4+}$. The measurements were obtained by the merged-beams technique using synchrotron radiation and the calculations were performed in the Dirac-Coulomb R-matrix approximation. Very satisfactory agreement was obtained, in fact, the best to date between measured and calculated photoionization cross sections along the tungsten isonuclear sequence. Improvement of the theoretical description had been expected with the charge state of the ion increasing and the experimental and theoretical work up to W$^{4+}$ demonstrate the validity of that assumption.

\ack

AM acknowledges support by Deutsche Forschungsgemeinschaft
under project numbers Mu-1068/20 and Mu-1068/22. RAP acknowledges a grant from the US Department
of Energy (DOE) under contract DE-FG02-03ER15424.
BMMcL acknowledges support by the US National Science Foundation through a grant to ITAMP
at the Harvard-Smithsonian Center for Astrophysics and Queen's University Belfast
for a visiting research fellowship (VRF).
The computational work was carried out at the National Energy Research Scientific
Computing Center in Berkeley, CA, USA and at the High Performance
Computing Center Stuttgart (HLRS) of the University of Stuttgart, Stuttgart, Germany.
The Advanced Light Source  is supported by the Director, Office of Science, Office of Basic Energy Sciences,
of the US Department of Energy under Contract No. DE-AC02-05CH11231.
%
%
%
%
\section*{References}

\begin{thebibliography}{10}
\expandafter\ifx\csname url\endcsname\relax
  \def\url#1{{\tt #1}}\fi
\expandafter\ifx\csname urlprefix\endcsname\relax\def\urlprefix{URL }\fi
\providecommand{\eprint}[2][]{\url{#2}}

\bibitem{Neu2013}
Neu R, Arnoux G, Beurskens M, Bobkov V, Brezinsek S, Bucalossi J, Calabro G,
  Challis C, Coenen J~W, {de la L}una E, {de V}ries P~C, Dux R, Frassinetti L,
  Giroud C, Groth M, Hobirk J, Joffrin E, Lang P, Lehnen M, Lerche E, Loarer T,
  Lomas P, Maddison G, Maggi C, Matthews G, Marsen S, Mayoral M~L, Meigs A,
  Mertens P, Nunes I, Philipps V, P\"{u}tterich T, Rimini F, Sertoli M, Sieglin
  B, Sips A~C~C, {van E}ester D, van Rooij G and {JET-EFDA Contributors} 2013
  {\em Phys. Plasmas\/} {\bf 20} 056111

\bibitem{Neu2003}
Neu R, Dux R, Geier A, Gruber O, Kallenbach A, Krieger K, Maier H, Pugno R,
  Rohde V, Schweizer S and {ASDEX Upgrade Team} 2003 {\em Fusion Eng. Des.\/}
  {\bf 65} 367 -- 374

\bibitem{Puetterich2010a}
P\"{u}tterich T, Neu R, Dux R, Whiteford A~D, {O}'{Mullane} M~G, Summers H~P
  and the {ASDEX}~{U}pgrade {T}eam 2010 {\em Nucl. Fusion\/} {\bf 50} 025012

\bibitem{Mueller2015b}
M\"{u}ller A 2015 {\em Atoms\/} {\bf 3} 120--161

\bibitem{Schippers2011b}
Schippers S, Bernhardt D, M\"{u}ller A, Krantz C, Grieser M, Repnow R, Wolf A,
  Lestinsky M, Hahn M, Novotn\'{y} O and Savin D~W 2011 {\em Phys. Rev. A\/}
  {\bf 83} 012711

\bibitem{Spruck2014}
Spruck K, Badnell N~R, Krantz C, Novotn\'{y} O, Becker A, Bernhardt D, Grieser
  M, Hahn M, Repnow R, Savin D~W, Wolf A, M\"{u}ller A and Schippers S 2014
  {\em Phys. Rev. A\/} {\bf 90} 032715

\bibitem{Badnell2016}
Badnell N~R, Spruck K, Krantz C, Novotn\'{y} O, Becker A, Bernhardt D, Grieser
  M, Hahn M, Repnow R, Savin D~W, Wolf A, M\"{u}ller A and Schippers S 2016
  {\em Phys. Rev. A\/} {\bf 93} 052703

\bibitem{Krantz2014}
Krantz C, Spruck K, Badnell N~R, Becker A, Bernhardt D, Grieser M, Hahn M,
  Novotn\'{y} O, Repnow R, Savin D~W, Wolf A, M\"{u}ller A and Schippers S 2014
  {\em J. Phys. Conf. Ser.\/} {\bf 488} 012051

\bibitem{Krantz2017}
Krantz C, Badnell N~R, M\"{u}ller A, Schippers S and Wolf A 2016 {\em J. Phys.
  B: At. Mol. Opt. Phys\/} {\bf 50} 052001

\bibitem{Rausch2011a}
Rausch J, Becker A, Spruck K, Hellhund J, {Borovik Jr} A, Huber K, Schippers S
  and M\"uller A 2011 {\em J. Phys. B: At. Mol. Opt. Phys\/} {\bf 44} 165202

\bibitem{Borovik2016}
{Borovik Jr} A, Ebinger B, Schury D, Schippers S and M\"{u}ller A 2016 {\em
  Phys. Rev. A\/} {\bf 93} 012708

\bibitem{Mueller2011a}
M\"{u}ller A, Schippers S, Kilcoyne A~L~D and Esteves D 2011 {\em Phys. Scr.\/}
  {\bf T144} 014052

\bibitem{Mueller2012}
M\"{u}ller A, Schippers S, Kilcoyne A~L~D, Aguilar A, Esteves D and Phaneuf R~A
  2012 {\em J. Phys. Conf. Ser.\/} {\bf 388} 022037

\bibitem{Mueller2014c}
M\"{u}ller A, Schippers S, Hellhund J, Kilcoyne A~L~D, Phaneuf R~A, Ballance
  C~P and McLaughlin B~M 2014 {\em J. Phys. Conf. Ser.\/} {\bf 488} 022032

\bibitem{Mueller2015h}
M\"{u}ller A, Schippers S, Hellhund J, Holste K, Kilcoyne A~L~D, Phaneuf R~A,
  Ballance C~P and McLaughlin B~M 2015 {\em J. Phys. B: At. Mol. Opt. Phys\/}
  {\bf 48} 235203

\bibitem{McLaughlin2016a}
McLaughlin B~M, Ballance C~P, Schippers S, Hellhund J, Kilcoyne A~L~D, Phaneuf
  R~A and M\"{u}ller A 2016 {\em J. Phys. B: At. Mol. Opt. Phys\/} {\bf 49}
  065201

\bibitem{Ballance2015a}
Ballance C and McLaughlin B~M 2015 {\em J. Phys. B: At. Mol. Opt. Phys.\/} {\bf
  48} 085201

\bibitem{Burke2011}
Burke P~G 2011 {\em {R-Matrix Theory of Atomic Collisions: Application to
  Atomic, Molecular and Optical Processes}\/} (Heidelberg Dordrecht London New
  York: Springer)

\bibitem{darc}
{DARC codes} \urlprefix\url{http://connorb.freeshell.org}

\bibitem{Beiersdorfer2015}
Beiersdorfer P, Clementson J and Safronova U~I 2015 {\em Atoms\/} {\bf 3} 260
  -- 272

\bibitem{Brown2014}
Brown B~T, Smirnov R~D and Krasheninnikov S~I 2014 {\em Phys. Plasmas\/} {\bf
  21} 024501

\bibitem{Phaneuf1999}
Phaneuf R~A, Havener C~C, Dunn G~H and M{\"u}ller A 1999 {\em Rep. Prog.
  Phys.\/} {\bf 62} 1143--1180

\bibitem{Schippers2016}
Schippers S, Kilcoyne A~L~D, Phaneuf R~A and M\"uller A 2016 {\em Contemp.
  Phys.\/} {\bf 57} 215--229

\bibitem{Covington2002a}
Covington A~M, Aguilar A, Covington I~R, Gharaibeh M~F, Hinojosa G, Shirley
  C~A, Phaneuf R~A, {\'A}lvarez I, Cisneros C, Dominguez-Lopez I, Sant'Anna
  M~M, Schlachter A~S, McLaughlin B~M and Dalgarno A 2002 {\em Phys. Rev. A\/}
  {\bf 66} 062710

\bibitem{Macaluso2016}
Macaluso D~A, Bogolub K, Johnson A, Aguilar A, Kilcoyne A~L~D, Bilodeau R~C,
  Bautista M, Kerlin A~B and Sterling N~C 2016 {\em J. Phys. B: At. Mol. Opt.
  Phys\/} {\bf 49} 235002

\bibitem{Trassl1997a}
Trassl R, Hathiramani P, Broetz F, Greenwood J~B, McCullough R~W, Schlapp M and
  Salzborn E 1997 {\em Phys. Scr.\/} {\bf T73} 380--381

\bibitem{Fricke1980a}
Fricke J, M{\"u}ller A and Salzborn E 1980 {\em Nucl. Instrum. Methods\/} {\bf
  175} 379--384

\bibitem{Rinn1982}
Rinn K, M{\"u}ller A, Eichenauer H and Salzborn E 1982 {\em Rev. Sci.
  Instrum.\/} {\bf 53} 829--837

\bibitem{Mueller2014b}
M\"{u}ller A, Schippers S, {Esteves-Macaluso} D, Habibi M, Aguilar A, Kilcoyne
  A~L~D, Phaneuf R~A, Ballance C~P and McLaughlin B~M 2014 {\em J. Phys. B: At.
  Mol. Opt. Phys.\/} {\bf 47} 215202

\bibitem{NIST2015}
Kramida A~E, Ralchenko Y, Reader J and {NIST ASD Team} 2017 {NIST Atomic
  Spectra Database (ver. 5.3)} online available from
  {http://physics.nist.gov/asd} \urlprefix\url{{http://physics.nist.gov/asd}}

\bibitem{Mueller2008a}
M{\"u}ller A 2008 {\em Adv. At. Mol. Opt. Phys.\/} {\bf 55} 293--417

\bibitem{Gumberidze2010}
Gumberidze A, Trassinelli M, Adrouche N, Szabo C~I, Indelicato P, Haranger F,
  Isac J~M, Lamour E, {Le Bigot} E~O, M\'{e}rot J, Prigent C, Rozet J~P and
  Vernhet D 2010 {\em Rev. Sci. Instrum.\/} {\bf 81} 033303

\bibitem{Hitz2000}
Hitz D, Melin G and Girard A 2000 {\em Rev. Sci. Instrum.\/} {\bf 71} 839 --
  845

\bibitem{Ballance2006}
Ballance C~P and Griffin D~C 2006 {\em J. Phys. B: At. Mol. Opt. Phys.\/} {\bf
  39} 3617

\bibitem{Fivet2012}
Fivet V, Bautista M~A and Ballance C~P 2012 {\em J. Phys. B: At. Mol. Opt.
  Phys.\/} {\bf 45} 035201

\bibitem{McLaughlin2012a}
McLaughlin B~M and Ballance C~P 2012 {\em J. Phys. B: At. Mol. Opt. Phys.\/}
  {\bf 45} 085701

\bibitem{McLaughlin2012b}
McLaughlin B~M and Ballance C~P 2012 {\em J. Phys. B: At. Mol. Opt. Phys.\/}
  {\bf 45} 095202

\bibitem{Norrington1987}
Norrington P~H and Grant I~P 1987 {\em J. Phys. B: At. Mol. Phys.\/} {\bf 20}
  4869-- 4881

\bibitem{Wijesundera1991}
Wijesundera W~P, Parpia F~A, Grant I~P and Norrington P~H 1991 {\em J. Phys. B:
  At. Mol. Opt. Phys.\/} {\bf 24} 1803

\bibitem{McLaughlin2015a}
McLaughlin B~M and Ballance C~P 2015 Petascale computations for large-scale
  atomic and molecular collisions {\em {Sustained Simulated Performance 2014}
  {Proceedings of the joint Workshop on Sustained Simulation Performance,
  University of Stuttgart (HLRS) and Tohoku University, 2014}\/} ed Resch M~M,
  Kovalenko Y, Fotch E, Bez W and Kobaysahi H (Cham Heidelberg New York
  Dordrecht London: Springer International Publishing) pp 173 -- 185

\bibitem{McLaughlin2015b}
McLaughlin B~M, Ballance C~P, Pindzola M~S and M\"{u}ller A 2015 {PAMOP:
  petascale atomic, molecular and optical collisions} {\em High Performance
  Computing in Science and Engineering '14\/} ed Nagel W~E, Kr\"{o}ner D~H and
  Resch M~M (Cham Heidelberg New York Dordrecht London: Springer International
  Publishing) pp 47 -- 61

\bibitem{McLaughlin2016c}
McLaughlin B~M, Ballance C~P, Pindzola M~S, Schippers S and M\"{u}ller A 2016
  {PAMOP Project: Petaflop Computations in Support of Experiments} {\em {High
  Performance Computing in Science and Engineering '15} {Transactions of the
  High Performance Computing Center, Stuttgart (HLRS) 2015}\/} ed Nagel W~E,
  Kr\"{o}ner D~H and Resch M~M (Cham Heidelberg New York Dordrecht London:
  Springer International Publishing) pp 51 -- 74

\bibitem{McLaughlin2017b}
McLaughlin B~M, Ballance C~P, Pindzola M~S, Stancil P~C, Schippers S and
  M\"{u}ller A 2017 {PAMOP Project: Computations in Support of Experiments and
  Astrophysical Applications} {\em {High Performance Computing in Science and
  Engineering '16} {Transactions of the High Performance Computing Center,
  Stuttgart (HLRS) 2016}\/} ed Nagel W~E, Kr\"{o}ner D~H and Resch M~M (Cham
  Heidelberg New York Dordrecht London: Springer International Publishing) pp
  33 -- 48

\bibitem{Macaluso2015}
Macaluso D~A, Aguilar A, Kilcoyne A~L~D, Red E~C, Bilodeau R~C, Phaneuf R~A,
  Sterling N~C and McLaughlin B~M 2015 {\em Phys. Rev. A\/} {\bf 92} 063424

\bibitem{Hinojosa2012}
Hinojosa G, Covington A~M, Alna'Washi G~A, Lu M, Phaneuf R~A, Sant'Anna M~M,
  Cisneros C, \'{A}lvarez I, Aguilar A, Kilcoyne A~L~D, Schlachter A~S,
  Ballance C~P and McLaughlin B~M 2012 {\em Phys. Rev. A\/} {\bf 86} 063402

\bibitem{Kennedy2014}
Kennedy E~T, Mosnier J~P, Kampen P~V, Cubaynes D, Guilbaud S, Blancard C,
  McLaughlin B~M and Bizau J~M 2014 {\em Phys. Rev. A\/} {\bf 90} 063409

\bibitem{Tyndall2016a}
Tyndall N~B, Ramsbottom C~A, Ballance C~P and Hibbert A 2016 {\em Mon. Not. R.
  Astron. Soc.\/} {\bf 456} 366 -- 373

\bibitem{Tyndall2016b}
Tyndall N~B, Ramsbottom C~A, Ballance C~P and Hibbert A 2016 {\em Mon. Not. R.
  Astron. Soc.\/} {\bf 462} 3350 -- 3360

\bibitem{Barthel2015}
Barthel M, Flesch R, R\"{u}hl E and McLaughlin B~M 2015 {\em Phys. Rev. A\/}
  {\bf 91} 013406

\bibitem{McLaughlin2017c}
McLaughlin B~M 2017 {\em Mon. Not. R. Astron. Soc.\/} {\bf 464} 1990 -- 1999

\bibitem{EnzongaYoca2012a}
{Enzonga Yoca} S, Palmeri P, Quinet P, Jumet G and Bi\'emont E 2012 {\em J.
  Phys. B: At. Mol. Opt. Phys\/} {\bf 45} 035002

\bibitem{Cowan1981}
Cowan R~D 1981 {\em The Theory of Atomic Structure and Spectra\/} (Berkeley:
  University of California Press)

\bibitem{Mueller2010b}
M\"{u}ller A, Schippers S, Phaneuf R~A, Kilcoyne A~L~D, Br\"{a}uning H,
  Schlachter A~S, Lu M and McLaughlin B~M 2010 {\em J. Phys. B: At. Mol. Opt.
  Phys\/} {\bf 43} 225201

\bibitem{Flambaum2015}
Flambaum V~V, Kozlov M~G and Gribakin G~F 2015 {\em Phys. Rev. A\/} {\bf 91}
  052704

\bibitem{Berengut2015}
Berengut J~C, Harabati C, Dzuba V~A, Flambaum V~V and Gribakin G~F 2015 {\em
  Phys. Rev. A\/} {\bf 92} 062717

\end{thebibliography}

\providecommand{\newblock}{}

\end{document}